\documentclass[11pt,letter,subeqn,fleqn]{article}
\usepackage{amsmath, amssymb,amsthm,cite}
\usepackage[usenames,dvipsnames]{color}
\usepackage{graphicx,subfigure}
\usepackage{tabulary}
\usepackage{float}
\usepackage{gensymb}
\usepackage{authblk}
\usepackage{geometry}

\title{Global Optimization of the Mean First Passage Time for Narrow Capture Problems in Elliptic Domains}

\author[1]{Jason Gilbert \thanks{Corresponding Author. Electronic mail: jtg074@mail.usask.ca}}
\author[1]{Alexei Cheviakov\thanks{Alternative English spelling: Alexey Shevyakov. Electronic mail: alexei.cheviakov@usask.ca}}
\affil[1]{Department of Mathematics and Statistics, University of Saskatchewan}

\def\beq{\begin{equation}}
\def\eeq{\end{equation}}
\def\barr{\begin{array}{ll}}
\def\earr{\end{array}}

\def\max{\mathop{\hbox{\rm max}}}

\def\vec#1{{\boldsymbol{\rm #1}}} 

{\theoremstyle{definition}

}

\newcounter{tabnum}

\newcommand{\ps}{\ + \ }
\newcommand{\ms}{\ - \ }
\newcommand{\es}{\ = \ }

\newcommand{\subDom}{\Omega_0}
\newcommand{\amfpt}{\overline{u}_0} 
\newcommand{\veps}{\varepsilon}     
\newcommand{\greenMat}{\mathcal{G}} 
\newcommand{\vecA}{\mathcal{A}}     
\newcommand{\vecE}{\textbf{e}}      
\newcommand{\trapLoc}{x}   
\newcommand{\domMeas}{|\Omega|}     
\newcommand{\etaS}{\eta_\star}      
\newcommand{\meritFunc}{q}          
\newcommand{\ecc}{\kappa}
\newcommand{\amfptRef}{\tilde{u}_0}

\newcommand{\wardTab}{2}

\begin{document}

\maketitle \numberwithin{equation}{section}
\maketitle \numberwithin{remark}{section}
\numberwithin{lemma}{section}
\numberwithin{proposition}{section}

\begin{abstract}

Narrow escape and narrow capture problems which describe the average times required to stop the motion of a randomly travelling particle within a domain have applications in various areas of science. While for general domains, it is known how the escape time decreases with the increase of the trap sizes, for some specific 2D and 3D domains, higher-order asymptotic formulas have been established, providing the dependence of the escape time on the sizes and locations of the traps. Such results allow the use of global optimization to seek trap arrangements that minimize average escape times. In a recent paper \cite{iyaniwura2021optimization}, an explicit size- and trap location-dependent expansion of the average mean first passage time (MFPT) in a 2D elliptic domain was derived. The goal of this work is to systematically seek global minima of MFPT for $1\leq N\leq 50$ traps in elliptic domains using global optimization techniques, and compare the corresponding putative optimal trap arrangements for different values of the domain eccentricity. Further, an asymptotic formula the for the average MFPT in elliptic domains with $N$ circular traps of arbitrary sizes is derived, and sample optimal configurations involving non-equal traps are computed.


\end{abstract}

\section{Introduction}

\begin{figure}[htbp]
\centering
\subfigure[]{\includegraphics[width=0.3\textwidth]{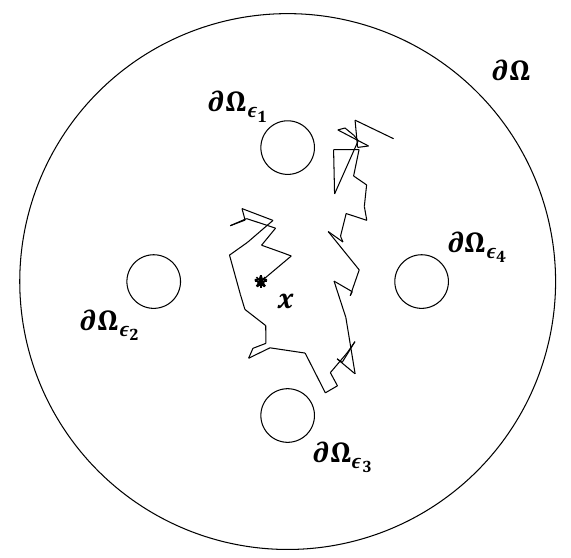}}
\hspace{5ex}
\subfigure[]{\includegraphics[width=0.3\textwidth]{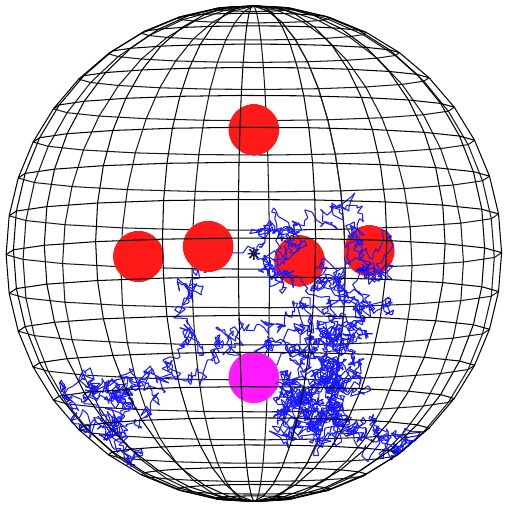}}
\caption{\small  (a) A two-dimensional narrow capture problem in the unit disk containing internal traps with absorbing boundaries $\{\partial\Omega_{\epsilon_j}\}$. (b) A three-dimensional narrow capture problem, a sample Brownian particle trajectory, leading to a capture in a trap (lowermost) denoted by purple color (color online).}
\label{fig:ProblemSchematic}
\end{figure}

The narrow capture problem, as described here, concerns the average time required for a particle undergoing Brownian motion to encounter a region within the domain, referred to as a trap, which causes its motion to cease. It is mathematically defined as a Neumann-Dirichlet Poisson problem
\begin{equation} \label{eq:defPDE}
\begin{split}
& \Delta u = -\dfrac{1}{D} \ , \quad x \in \subDom \ ; \qquad
\subDom = \Omega \setminus \mathop{\cup}_{j = 1}^{N} \Omega_{\veps_j} \ ; \\[5pt]
& \partial_n u = 0 \ , \quad x \in \partial\Omega \ ; \qquad
u = 0 \ , \quad x \in \partial\Omega_{\veps_j} \ , \quad j = 1, \ldots , N \
\end{split}
\end{equation}
where $\Omega\subset \mathbb{R}^n$, $n=2,3$, denotes the physical domain of the problem; $\{\Omega_{\veps_j}\}_{j = 1}^{N}$ are small trap domains within $\Omega$, $\subDom$ is the domain except the traps, where the motion of particles takes place, and $\partial_n$ is the normal derivative on the domain boundary $\partial\Omega$. The diffusivity $D$ of the medium filling $\subDom$ is assumed constant. The problem  \eqref{eq:defPDE} describes the distribution of the mean capture time, the time $u(x)$ needed for a  particle to be captured by any trap, averaged over a large number of launches from the same point $x\in \subDom$. An illustration of the problem is provided in Figure \ref{fig:ProblemSchematic}.

The boundary conditions on $\partial \Omega$ are reflective: $\partial_n u = 0$, whereas the traps $\Omega_{\veps_j}$ are characterized by immediate absorption of a Brownian particle, which is manifested by a Dirichlet boundary conditions $u = 0$ on all $\partial\Omega_{\veps_j}$.

The above generic problem \eqref{eq:defPDE} affords a variety of physical interpretations, ranging from biological to electrostatic (see, e.g., Refs.~\cite{red2001, holcman2014narrow} for an overview of applications). In this work, it will be strictly considered in terms of a particle undergoing Brownian motion \cite{saffman1975brownian}. In this case, the problem regards the stopping time
\cite{red2001, ralf2014first} of a Brownian particle. When the path of the particle intersects the boundary of one of the traps, the particle is captured. This capture time may be interpreted as the time required for the particle to leave the confining domain, thus it is often referred to as the first passage time \cite{bressloff2008diffusion, bressloff2013stochastic, holcman2004escape}. As Brownian motion is an inherently random process, the mean first passage time (MFPT) is considered. Interpreting the problem \eqref{eq:defPDE} accordingly, $u$ is the MFPT; $D$ is the diffusion coefficient, representing the mobility of the Brownian particle;  $\Omega_{\veps_j}$ is the portion of the domain occupied by the $j_{th}$ trap.

Given the physical domain and the number and sizes of the traps, it is of interest to ask whether there is an \emph{optimal} arrangement of $N$ traps within the domain which minimizes the MFPT, or in other words, maximizes the rate at which Brownian particles encounter the absorbing traps. Related work dedicated to similar optimization, in the case that the traps are asymptotically small relative to the size of the domain, for various kinds of confining domains with interior or boundary traps, can be found, for example, in Refs.~\cite{kolokolnikov2005optimizing, iyaniwura2021optimization, cheviakov2011optimizing, gilbert2019globally, ridgway2019locally} and references therein. Both putative globally optimal and multiple locally optimal arrangements of boundary and volume traps have been computed in various settings. An important aspect of such computations is the existence of a large number of locally optimal particle arrangements with very close merit function values. This number quickly grows with $N$, increasing the computational difficulty of the determination of the globally optimal configuration; see, e.g.,  Ref.~\cite{ridgway2019locally} where locally optimal configurations were systematically computed for particles located on the unit sphere, and the number of local minima exhibits exponential growth as $N$ increases.

In the current contribution, we consider the narrow capture problem for the case of a 2D elliptical domain, elaborating on previous work on the subject for the case of a circular domain \cite{kolokolnikov2005optimizing}, and the more general case of the elliptical domain \cite{iyaniwura2021optimization}, and examine specifically the case where traps are not of equal size. The paper is organized as follows. In Section \ref{sec:problem} we briefly summarise results this work is based on. This includes the approximate asymptotic MFPT formulas that hold in the case that the traps are small relative to the domain size and are well-separated, as well as the choice of merit function for average MFPT (AMFPT) optimization.

Section \ref{sec:method} describes the optimization method chosen in the current study, the algorithms, the details of their use, and some decisions made to facilitate comparison to previous studies. Specifically, we seek the optimal configuration of traps for numbers of traps $N \leq 50$, and elliptic domains of sample eccentricities $0$, $0.472$, $0.802$, and $0.995$. In Section \ref{sec:results} the results of the study for $N$ identical traps are presented. These results include the optimized values of the merit functions (related to the domain-average Brownian particle capture time) for each number of traps, and each domain eccentricity; in the case of the unit disk, the new computations are compared to those of the previous study \cite{kolokolnikov2005optimizing}. Distributions of traps for some illustrative cases are shown, and bulk measures of trap distribution including the closest-neighbor pairwise distance, smallest distance to the boundary, and area per trap are calculated for each of the optimized configurations of identical traps. It is also shown that unlike the cases of the disk- and sphere-shaped domains \cite{kolokolnikov2005optimizing, gilbert2019globally}, optimal trap locations in ellipses are not generally accumulated on scaled versions of the domain boundary.

Section \ref{sec:traps:different} is concerned with a more general case of unequal traps in an elliptic domain. Asymptotic formulas for the AMFPT are derived for this case, generalizing those for the case of identical traps. Sample putative optimal configurations for traps of two kinds, for different trap size relations and domain eccentricities, are computed.

Section \ref{sec:discussion} contains a discussion of the results and some related problems.

\section{Asymptotic MFPT for the elliptic domain with $N$ identical traps} \label{sec:problem}

The main goal of this contribution is to further explore optimal configurations of absorbing traps found using asymptotic expansions for the circular and elliptical domains for which asymptotic MFPT formulae depending on trap positions are now available \cite{kolokolnikov2005optimizing, iyaniwura2021optimization}. To this end, numerical methods will be used to approximate the optimum configurations of larger numbers of traps than have previously been considered. In the case of the unit disk, the parameter space used in this study is general and does not assume any simplifying restrictions that have been imposed in previous works to reduce computational complexity. To begin, we recall the formulas for identical traps \cite{iyaniwura2021optimization}, following which, in Section \ref{sec:traps:different}, we derive the corresponding formulas for traps of differing sizes.

In essence, the problem at hand is to find the trap positions which minimize the spatial average of the MFPT $u(x)$ in the elliptic domain $\Omega$ of area $\domMeas$, denoted by
\[
\bar{u}=\dfrac{1}{\domMeas}\int u(x)\,dS
\]
and approximated by $\amfpt$ in the leading order in terms of the deviation $\sigma$ of the domain from the unit disk \cite{iyaniwura2021optimization}. We now summarize the equations used to minimize $\amfpt$, as derived in \cite{kolokolnikov2005optimizing, iyaniwura2021optimization}. The unit disk will be considered a special case of the elliptical domain with zero eccentricity. In either case, when the domain contains $N$ identical circular traps, each of radius $\veps$, the approximate AMFPT satisfies \cite{iyaniwura2021optimization}
\begin{equation} \label{eq:ellAMFPT}
\amfpt \es \dfrac{\domMeas}{2\pi D\nu N} \ps \dfrac{2\pi}{N}\vecE^{T}\greenMat \vecA \ ,
\end{equation}
where the column vector $\vecA$ satisfies
\begin{equation} \label{eq:ellAMFPT:A:sum}
\sum_{j=1}^{N} \vecA_j = \dfrac{\domMeas}{2\pi D}
\end{equation}
and is given by the solution of the linear system
\begin{equation} \label{eq:ellAMFPT:A}
\left[ I + 2\pi\nu\left( I - E \right)\greenMat \right]\vecA \es \dfrac{\domMeas}{2\pi D N}\vecE \ .
\end{equation}
Here $E \equiv \vecE\vecE^{T}/N$, $\vecE = (1, \ldots, 1)^{T}$, $\nu \equiv -1/\log\veps$, and the Green's matrix $\greenMat$ depends on the trap center locations $\lbrace \trapLoc_1, \ldots , \trapLoc_N \rbrace$, such that
\begin{equation} \label{eq:ellGmat}
\greenMat_{ij} \es \left\lbrace
\begin{array}{ll}
G(\trapLoc_i ; \trapLoc_j), & i \neq j, \\[5pt]
R_i,  & i = j.
\end{array}
\right.
\end{equation}
In \eqref{eq:ellGmat}, $G(\trapLoc_i ; \trapLoc_j)$ is the Green's function of the domain, and $R_i\equiv R(\trapLoc_i)$ is the regular part of $G(\trapLoc_i ; \trapLoc_j)$ as $\trapLoc_j\to \trapLoc_i$.

Examining the equation \eqref{eq:ellAMFPT}, it can be seen that the first term depends only on the combined trap size but does not depend on the trap locations. The second term explicitly depends on the trap locations is the quantity and therefore can be optimized by adjusting trap positions. The merit function subject to optimization can be chosen as
\begin{equation} \label{eq:ellMerit}
\meritFunc\left( \trapLoc_1,\ldots,\trapLoc_N \right) \es \vecE^{T}\greenMat \vecA \
\end{equation}
depending on $2N$ coordinates of $N$ traps located within the elliptical domain. For a value of $\meritFunc$ to be a permissible solution, it is required that $\amfpt \geq 0$, as it is a measure of time; all traps must be within the domain; no trap may be in contact with any other trap (or the two must instead be represented as a single non-circular trap).

While the preceding statements are true for both the circular and the elliptical domain, the elements of the matrix $\greenMat$ are populated by evaluating the Green's function of the domain for each pair of traps, and the form of this function differs for the two cases considered here.

In the case of the circular domain, the elements of the Green's matrix $\greenMat$ are computed using the Neumann Green's function \cite{kolokolnikov2005optimizing}
\begin{subequations}
\begin{equation} \label{cirGreenFunc}
G(\trapLoc ; \trapLoc_0) \es \dfrac{1}{2\pi}\left(
-\log|\trapLoc - \trapLoc_0| \ms \log\left| \trapLoc|\trapLoc_0| - \dfrac{\trapLoc_0}{|\trapLoc_0|} \right| \ps \dfrac{1}{2}\left(|\trapLoc|^2 + |\trapLoc_0|^2\right) \ms \dfrac{3}{4}
\right) \ ,
\end{equation}
and its regular part
\begin{equation}
R(\trapLoc) \es \dfrac{1}{2\pi}\left(
-\log\left| \trapLoc|\trapLoc| - \dfrac{\trapLoc}{|\trapLoc|} \right| \ps |\trapLoc|^2 \ms \dfrac{3}{4}
\right) \ .
\end{equation}
\end{subequations}

\medskip The Green's function for the elliptical domain, in the form of a quickly convergent series, has been derived in Ref.~\cite{iyaniwura2021optimization} using elliptical coordinates $(\xi,\eta)$. It has the form
\begin{equation} \label{eq:ellGreenFunc}
\begin{split}
G(\trapLoc, \trapLoc_0) \es \dfrac{1}{4\domMeas}\left( |\trapLoc|^2 + |\trapLoc_0|^2 \right) \ms \dfrac{3}{16\domMeas}(a^2 + b^2) \ms \dfrac{1}{4\pi}\log\beta \ms \dfrac{1}{2\pi}\xi_>  \\[5pt]
\ms \dfrac{1}{2\pi}\sum^{\infty}_{n=0}\log\left( \prod_{j=1}^{8} |1 - \beta^{2n}z_j| \right) \ ,
\qquad
\trapLoc \neq \trapLoc_0 \ ,
\end{split}
\end{equation}
where $a$ and $b$ are the major and minor axis of the domain, respectively; the area of the domain is $\domMeas = \pi ab$, the parameter $\beta = (a - b)/(a + b)$, and the values $\xi_> = \max(\xi, \xi_0)$, $z_1, \ldots, z_8$ are defined in terms of the elliptical coordinates $\xi$ and $\eta$ as follows.

The Cartesian coordinates $(x,y)$ and the elliptical coordinates $(\xi,\eta)$ are related by the transformation
\begin{subequations}
\begin{equation} \label{eq:ellCoordsA}
x = f\cosh\xi\cos\eta \ , \quad
y = f\sinh\xi\sin\eta \ , \quad
f = \sqrt{a^2 - b^2} \ .
\end{equation}
For the major and minor axis of the elliptical domain, one has
\begin{equation} \label{eq:ellCoordsB}
a = f\cosh\xi_b \ , \quad
b = f\sinh\xi_b \ , \quad
\xi_b = \tanh^{-1}\left(\dfrac{b}{a}\right) = -\dfrac{1}{2}\log\beta \ .
\end{equation}
\end{subequations}
For the backward transformation, to determine $\xi(x, y)$, equation \eqref{eq:ellCoordsA} is solved to give
\begin{subequations}
\begin{equation}
\xi = \dfrac{1}{2}\log\left( 1 - 2s + 2\sqrt{s^2 - s} \right) \ , \quad
s \equiv \dfrac{-\mu - \sqrt{\mu^2 + 4f^2y^2}}{2f^2} \ , \quad
\mu \equiv x^2 + y^2 - f^2 \ .
\end{equation}
In a similar fashion, $\eta(x, y)$ is found in terms of $\etaS \equiv \sin^{-1}(\sqrt{p})$ to be
\begin{equation}
\eta \es \left\lbrace \begin{array}{ll}
\etaS \ ,        & \text{if}\ x \geq 0,\ y \geq 0 \\[5pt]
\pi - \etaS \ ,  & \text{if}\ x < 0,\ y \geq 0 \\[5pt]
\pi + \etaS \ ,  & \text{if}\ x \leq 0,\ y < 0 \\[5pt]
2\pi - \etaS \ , & \text{if}\ x > 0,\ y < 0 \\[5pt]
\end{array}
\right. \ ,
\qquad
p\equiv \dfrac{-\mu + \sqrt{\mu^2 + 4f^2y^2}}{2f^2} \ .
\end{equation}
\end{subequations}
Finally, the $z_j$-terms appearing in the infinite sum of equation \eqref{eq:ellGreenFunc} are defined via $\xi$, $\eta$, and $\xi_b$ as
\begin{equation} \label{eq:ztermsG}
\begin{array}{lll}
z_1 \equiv e^{-|\xi - \xi_0| + i(\eta - \eta_0)} \ ,          &
z_2 \equiv e^{ |\xi - \xi_0| - 4\xi_b + i(\eta - \eta_0)} \ , &
z_3 \equiv e^{-(\xi + \xi_0) - 2\xi_b + i(\eta - \eta_0)} \ , \\[4pt]
z_4 \equiv e^{ (\xi + \xi_0) - 2\xi_b + i(\eta - \eta_0)} \ , &
z_5 \equiv e^{ (\xi + \xi_0) - 4\xi_b + i(\eta + \eta_0)} \ , &
z_6 \equiv e^{-(\xi + \xi_0) + i(\eta + \eta_0)}          \ , \\[4pt]
z_7 \equiv e^{|\xi - \xi_0| - 2\xi_b + i(\eta + \eta_0)}  \ , &
z_8 \equiv e^{-|\xi - \xi_0| - 2\xi_b + i(\eta + \eta_0)} \ . &
\\
\end{array}
\end{equation}

The regular part of the Neumann Green's function, $R$, can be expressed in similar terms as $G$ in equation \eqref{eq:ellGreenFunc} but requires a restatement of the $z_j$-terms given in \eqref{eq:ztermsG}. It is given by
\begin{subequations}
\begin{equation} \label{eq:ellGreenFuncR}
\begin{split}
R(\trapLoc_0) \es & \dfrac{|\trapLoc_0|^2}{2\domMeas} \ms \dfrac{3}{16\domMeas}(a^2 + b^2) \ps \dfrac{1}{2\pi}\log(a + b) \ms \dfrac{\xi_0}{2\pi} \ps \dfrac{1}{4\pi}\log\left( \cosh^2\xi_0 - \cos^2\eta_0 \right) \\[5pt]
& \ms \dfrac{1}{2\pi}\sum_{n=1}^{\infty}\log\left(1 - \beta^{2n}\right) \ms \dfrac{1}{2\pi}\sum_{n=0}^{\infty}\log\left( \prod_{j=2}^{8}\left|1 - \beta^{2n}z^{0}_j\right| \right).
\end{split} \
\end{equation}
Here $z^{0}_j$ denotes the limiting value of $z_j$, as defined in equation \eqref{eq:ztermsG}, as $(\xi, \eta) \to (\xi_0, \eta_0)$, given by
\begin{equation}
\begin{array}{lll}
                                             &
z^{0}_2 = \beta^2                        \ , &
z^{0}_3 = \beta e^{-2\xi_0}              \ , \\[5pt]
z^{0}_4 = \beta e^{2\xi_0}               \ , &
z^{0}_5 = \beta^2 e^{2(\xi_0 + i\eta_0)} \ , &
z^{0}_6 = e^{2(-\xi_0 + i\eta_0)}        \ , \\[5pt]
z^{0}_7 = \beta e^{2i\eta_0}             \ , &
z^{0}_8 = \beta e^{2i\eta_0}             \ . &
\\
\end{array}
\end{equation}
\end{subequations}

\section{Global optimization} \label{sec:method}

In this section, the methods used to find the optimum trap configurations minimizing the average MFPT \eqref{eq:ellAMFPT} are discussed. We include a description of the general strategy for optimization, the algorithms used, and their specific implementations.

For each problem the same general approach was taken to finding the optimum. This was to search iteratively, switching between global and local searches after each iteration. The global search used the particle swarm algorithm \texttt{PSwarm} \cite{vaz2007particle}, as implemented in the freely available software package OPTI \cite{currie2012opti}. The default values for the social and cognitive parameters were chosen, meaning the local optimum known by each particle tended to be as attractive as the known global optimum. These values were chosen with the intent that the parameter space would be explored as broadly as possible. For the local search, the Nelder-Mead algorithm \cite{lagarias1998convergence}, as implemented in MATLAB R2020, was used. The algorithms were chosen based on their generality and simplicity to adapt them to the current study.

In addition, for the case of the elliptical domain, special care needed to be taken to ensure that the traps were well-separated. If the traps were to come into contact, or overlap with one another, the asymptotic equation \eqref{eq:ellAMFPT} can yield non-physical values $\amfpt < 0$, (which is a common feature of asymptotic formulas that replace finite traps with ``point traps" in various narrow escape and narrow capture setups). In the MFPT optimization, the traps are effectively repelled from one another, as well as from their ``reflections" in the domain boundary; this is mathematically manifested in the fact that the Green's function \eqref{cirGreenFunc} for the disk domain grows logarithmically as $\trapLoc\to \trapLoc_0$, as well as when $|\trapLoc|\to 1$. In particular, $\meritFunc$ increases as distance between traps decreases, as traps begin to overlap, $\meritFunc$ decreases extremely rapidly, appearing to the optimization algorithm to be a favourable configuration. Though this problem can be addressed by artificially assigning $\meritFunc$ a very high value when an unacceptable configuration is encountered, this approach was found to significantly reduce the effectiveness of the global search, as many evaluations of $\meritFunc$ would be wasted on these configurations. Instead, an optimum was first found using the iterative approach taking $\veps = 0$, following which a local search was carried out using these coordinates as an initial guess for a subsequent search.

For traps of equal size we took $\veps = 0.05$ in order to facilitate comparison with previous studies \cite{iyaniwura2021optimization}. For traps of different sizes (see Section \ref{sec:traps:different} below) we took two traps to be larger than the others. The size of the smaller traps was chosen to be $\veps_1 = 10^{-9}$ to further reduce the number of computations wasted on insufficiently separated traps, and the larger traps were taken to be either $\veps_2 = 10^{3}\veps_1$ or $\veps_2 = 10^{6}\veps_1$ times larger. The relative difference was chosen to be several orders of magnitude because the trap interactions are weighted by $\nu = -\log(\veps)^{-1}$, meaning a significant different in sizes is required to achieve observable effects. The number of larger traps was kept small for the following reasons. For traps of different sizes it was found that the problem was greatly complicated by the fact that the optimal placement of a trap now depended on its size. This meant that some advantageous symmetry had been lost, in a similar way that changing the eccentricity of the domain eliminates the rotational invariance of a configuration. To account for this we used an optimized configuration obtained for identical traps as an initial guess, then ran subsequent optimizations for all unique combinations of initial trap locations. Consequently, for $N$ total traps and $m$ traps of different sizes, there are $N$ choose $m$ combinations to consider, which roughly grows like $N^{m}/m!$. Due to the computational complexity of the problem, we limited our study to $N = 5$, $N = 10$, and $m = 2$.

Defining the eccentricity of the elliptic domain according to the usual formula
\begin{equation} \label{eq:eccDef}
\ecc \es \sqrt{1 - \left( \dfrac{b}{a} \right)^2} \ ,
\end{equation}
where $a$ is the major axis and $b$ the minor, optimum configurations were computed for $N \leq 50$ for domain eccentricities of $\ecc = 0$, $0.472$, $0.802$, and $0.995$. For each eccentricity value, the axes of the ellipse were chosen so that the area of the domain was fixed at $\domMeas = \pi$, to allow for natural comparisons.

\section{Optimization results: $N$ identical traps} \label{sec:results}

In this section we use the method outlined in Section \ref{sec:method} to compute putative global optimum configurations of $N$ equal traps in the ellipse, $1\leq N\leq 50$, compare these results of this study to previous results for the unit disk \cite{kolokolnikov2005optimizing}, and present some analysis of the trap configurations. In particular, we show that the unit disk AMFPT values for the putative optimal configurations found below are consistently better than those given in Ref.~\cite{kolokolnikov2005optimizing}.

\begin{figure}[htbp]
\centering
\includegraphics[width=0.6\textwidth]{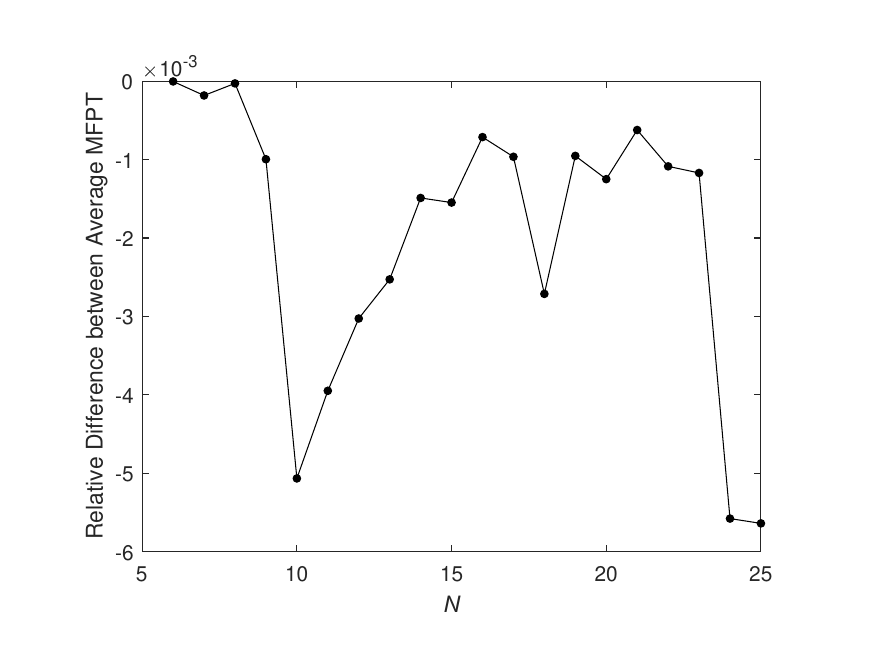}
\caption{\small Relative difference $(\amfpt - \amfptRef)/\amfptRef$ between average MFPT $\amfptRef$ in the unit disk for previously found putative optimal configurations (Ref.~ \cite{kolokolnikov2005optimizing}, Table \wardTab) and the optimal average MFPT values $\amfpt$ \eqref{eq:amfpt:new} computed in this work (Table \ref{tab:amfptComp}). }
\label{fig:amfptComp}
\end{figure}

Though it is conventional to discuss optimal configurations in terms of their ``interaction energy" used as the merit function (see, e.g., Ref.~\cite{ridgway2019locally} and references therein), here we will use the full asymptotic AMFPT expression \cite{iyaniwura2021optimization, kolokolnikov2005optimizing}
\begin{equation}\label{eq:amfpt:new}
\amfpt \es \dfrac{\domMeas}{2\pi D\nu N}\left( 1 + \dfrac{4\pi^2 D \nu}{\domMeas}\vecE^{T}\greenMat \vecA \right) \ .
\end{equation}
which facilitates comparisons with the previous work \cite{kolokolnikov2005optimizing} in the case of the unit disk. We denote the AMFPT values of the trap arrangements found in Ref.~\cite{kolokolnikov2005optimizing} by $\amfptRef$. Table \ref{tab:amfptComp} compares $\amfptRef$ and $\amfpt$ \eqref{eq:amfpt:new} for each $N$ reported in the previous study (Ref.~\cite{kolokolnikov2005optimizing}, Table \wardTab). It can be seen that the new values are consistently smaller, differing at most in the third significant figure. A plot of the difference between the previous and the new putative optimal AMFPT values, relative to the previous results, is presented in Figure \ref{fig:amfptComp}.

\begin{table}[htbp]
\begin{center}
\begin{tabular}{| c | c | l |}
\hline
$N$ & $\amfptRef$ & $\amfpt$ \\
6 & 0.11648 & 0.11648 \\
7 & 0.09299 & 0.09297 \\
8 & 0.07660 & 0.07660 \\
9 & 0.06518 & 0.06512 \\
10 & 0.05653 & 0.05624 \\
11 & 0.04920 & 0.04900 \\
12 & 0.04291 & 0.04278 \\
13 & 0.03805 & 0.03796 \\
14 & 0.03380 & 0.03375 \\
15 & 0.03042 & 0.03038 \\
16 & 0.02747 & 0.02745 \\
17 & 0.02502 & 0.02499 \\
18 & 0.02286 & 0.02280 \\
19 & 0.02078 & 0.02076 \\
20 & 0.01909 & 0.01907 \\
21 & 0.01756 & 0.01755 \\
22 & 0.01626 & 0.01624 \\
23 & 0.01512 & 0.01510 \\
24 & 0.01411 & 0.01403 \\
25 & 0.01314 & 0.01307 \\
\hline
\end{tabular}
\end{center}
\caption{\small Average MFPT $\amfptRef$ in the unit disk for previously computed putative optimal configurations (Ref.~ \cite{kolokolnikov2005optimizing}, Table \wardTab) compared to the AMFPT $\amfpt$ \eqref{eq:amfpt:new} for optimal trap arrangements computed in this work. The relative difference plot is given in Figure \ref{fig:amfptComp}.
}
\label{tab:amfptComp}
\end{table}

The computed optimal values of the merit function \eqref{eq:ellMerit} that correspond to putative globally optimal minima of the average MFPT \eqref{eq:ellAMFPT}, for the domain eccentricities $\ecc = 0$ (circular disk), $0.472$, $0.802$, and $0.995$, are presented in Table \ref{tab:meritTable} below, and are graphically shown in Figure \ref{fig:meritComp}. While the first three plots are nearly identical, the plot (d) for the largest eccentricity value differs significantly for small $N$, but becomes similar to the other plots for larger $N$.

\begin{figure}[H]
\centering
\includegraphics[width=0.6\textwidth]{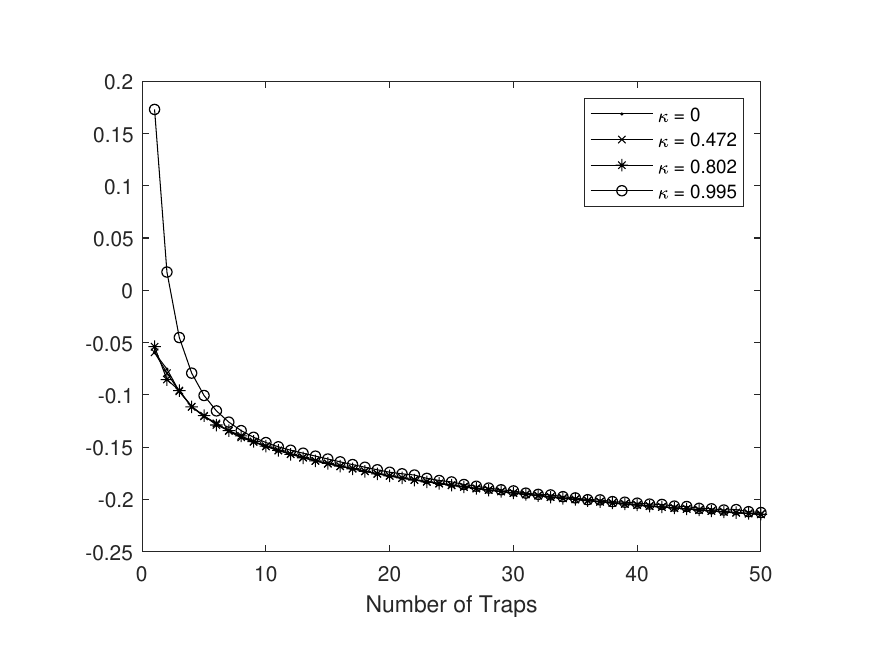}
\caption{\small The putative optimal values of the merit function \eqref{eq:ellMerit} for different ellipse eccentricity values as a function of the number of traps $N$ (Table \ref{tab:meritTable}).}
\label{fig:meritComp}
\end{figure}

\newgeometry{left=2cm,right=2cm,top=2cm,bottom=2cm}
\begin{table}[htbp]
\begin{center}
\begin{tabular}{| c | c | c | c | c |}
\hline
$N$    & \multicolumn{4}{c|}{Merit Value} \\
  & $\ecc = 0$ & $\ecc = 0.472$ & $\ecc = 0.802$ & $\ecc = 0.995$ \\
\hline
1 & -0.0597 & -0.0594 & -0.0540 & 0.1730 \\
2 & -0.0754 & -0.0792 & -0.0854 & 0.0175 \\
3 & -0.0969 & -0.0967 & -0.0959 & -0.0452 \\
4 & -0.1112 & -0.1113 & -0.1115 & -0.0793 \\
5 & -0.1207 & -0.1207 & -0.1200 & -0.1007 \\
6 & -0.1272 & -0.1274 & -0.1289 & -0.1154 \\
7 & -0.1348 & -0.1347 & -0.1342 & -0.1261 \\
8 & -0.1409 & -0.1408 & -0.1393 & -0.1343 \\
9 & -0.1451 & -0.1451 & -0.1447 & -0.1407 \\
10 & -0.1489 & -0.1494 & -0.1492 & -0.1457 \\
11 & -0.1526 & -0.1532 & -0.1533 & -0.1498 \\
12 & -0.1567 & -0.1566 & -0.1569 & -0.1530 \\
13 & -0.1599 & -0.1598 & -0.1603 & -0.1559 \\
14 & -0.1632 & -0.1632 & -0.1632 & -0.1587 \\
15 & -0.1659 & -0.1660 & -0.1657 & -0.1614 \\
16 & -0.1685 & -0.1686 & -0.1683 & -0.1642 \\
17 & -0.1708 & -0.1705 & -0.1708 & -0.1668 \\
18 & -0.1731 & -0.1731 & -0.1733 & -0.1693 \\
19 & -0.1756 & -0.1755 & -0.1756 & -0.1718 \\
20 & -0.1777 & -0.1776 & -0.1775 & -0.1741 \\
21 & -0.1798 & -0.1797 & -0.1796 & -0.1756 \\
22 & -0.1815 & -0.1815 & -0.1816 & -0.1768 \\
23 & -0.1831 & -0.1831 & -0.1833 & -0.1800 \\
24 & -0.1848 & -0.1851 & -0.1848 & -0.1820 \\
25 & -0.1864 & -0.1867 & -0.1864 & -0.1834 \\
\hline
\end{tabular}
\begin{tabular}{| c | c | c | c | c |}
\hline
$N$    & \multicolumn{4}{c|}{Merit Value} \\
  & $\ecc = 0$ & $\ecc = 0.472$ & $\ecc = 0.802$ & $\ecc = 0.995$ \\
\hline
26 & -0.1880 & -0.1882 & -0.1880 & -0.1858 \\
27 & -0.1897 & -0.1896 & -0.1896 & -0.1875 \\
28 & -0.1911 & -0.1910 & -0.1911 & -0.1893 \\
29 & -0.1925 & -0.1925 & -0.1925 & -0.1909 \\
30 & -0.1940 & -0.1940 & -0.1938 & -0.1920 \\
31 & -0.1953 & -0.1953 & -0.1952 & -0.1941 \\
32 & -0.1964 & -0.1964 & -0.1965 & -0.1953 \\
33 & -0.1977 & -0.1978 & -0.1978 & -0.1958 \\
34 & -0.1989 & -0.1989 & -0.1990 & -0.1975 \\
35 & -0.2000 & -0.2001 & -0.2000 & -0.1987 \\
36 & -0.2012 & -0.2012 & -0.2012 & -0.2003 \\
37 & -0.2025 & -0.2024 & -0.2023 & -0.2005 \\
38 & -0.2035 & -0.2035 & -0.2033 & -0.2022 \\
39 & -0.2045 & -0.2044 & -0.2043 & -0.2028 \\
40 & -0.2056 & -0.2055 & -0.2053 & -0.2037 \\
41 & -0.2065 & -0.2065 & -0.2064 & -0.2046 \\
42 & -0.2074 & -0.2073 & -0.2073 & -0.2049 \\
43 & -0.2083 & -0.2084 & -0.2083 & -0.2065 \\
44 & -0.2093 & -0.2092 & -0.2092 & -0.2069 \\
45 & -0.2102 & -0.2103 & -0.2102 & -0.2086 \\
46 & -0.2110 & -0.2111 & -0.2111 & -0.2091 \\
47 & -0.2119 & -0.2120 & -0.2119 & -0.2102 \\
48 & -0.2127 & -0.2128 & -0.2128 & -0.2098 \\
49 & -0.2136 & -0.2136 & -0.2136 & -0.2118 \\
50 & -0.2144 & -0.2143 & -0.2143 & -0.2126 \\
\hline
\end{tabular}
\end{center}
\caption{\small Optimized values of the merit function \eqref{eq:ellMerit}, for each number of traps $N$ and eccentricity $\ecc$ considered in this study. Plots of these values are found in Figure \ref{fig:meritComp}.}
\label{tab:meritTable}
\end{table}
\restoregeometry

\begin{figure}[htbp]
\begin{subfigure}[]{
\includegraphics[width=0.5\textwidth]{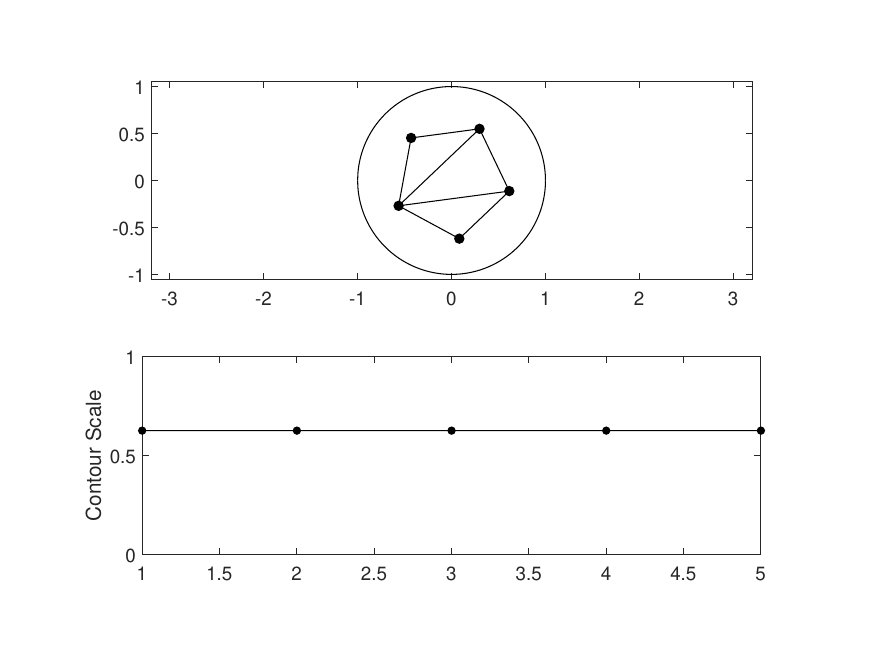}
}
\end{subfigure}
\begin{subfigure}[]{
\includegraphics[width=0.5\textwidth]{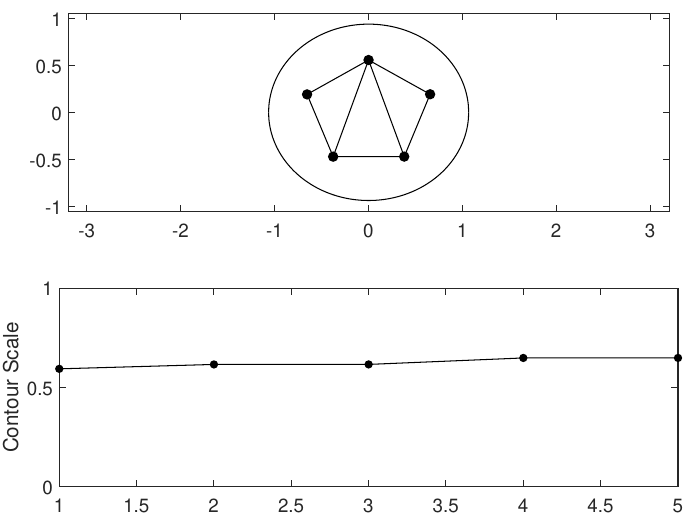}
}
\end{subfigure}
\begin{subfigure}[]{
\includegraphics[width=0.5\textwidth]{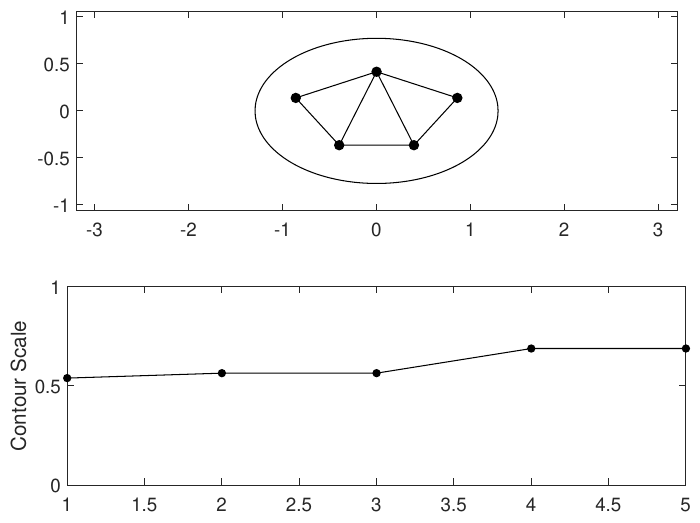}
}
\end{subfigure}
\begin{subfigure}[]{
\includegraphics[width=0.5\textwidth]{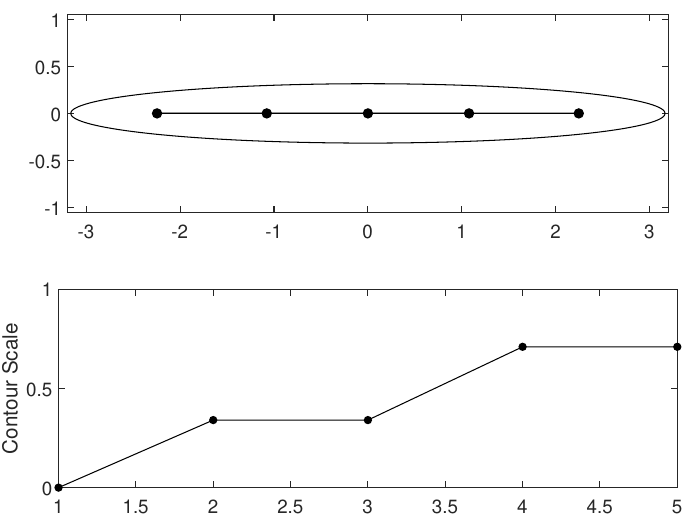}
}
\end{subfigure}
\caption{\small Plots depicting optimal trap distribution for $N=5$, comparing eccentricities of (a) $\ecc = 0$, (b) $\ecc = 0.472$, (c) $\ecc = 0.802$, (d) $\ecc = 0.995$. Upper plots show positions of traps along with a crude visualization of nearest-neighbour pairs calculated using Delaunay triangulation. Lower plots show the scaling factor given by the equation \eqref{eq:scaleFact}.}
\label{fig:configN5}
\end{figure}

\begin{figure}[htbp]
\begin{subfigure}[]{
\includegraphics[width=0.5\textwidth]{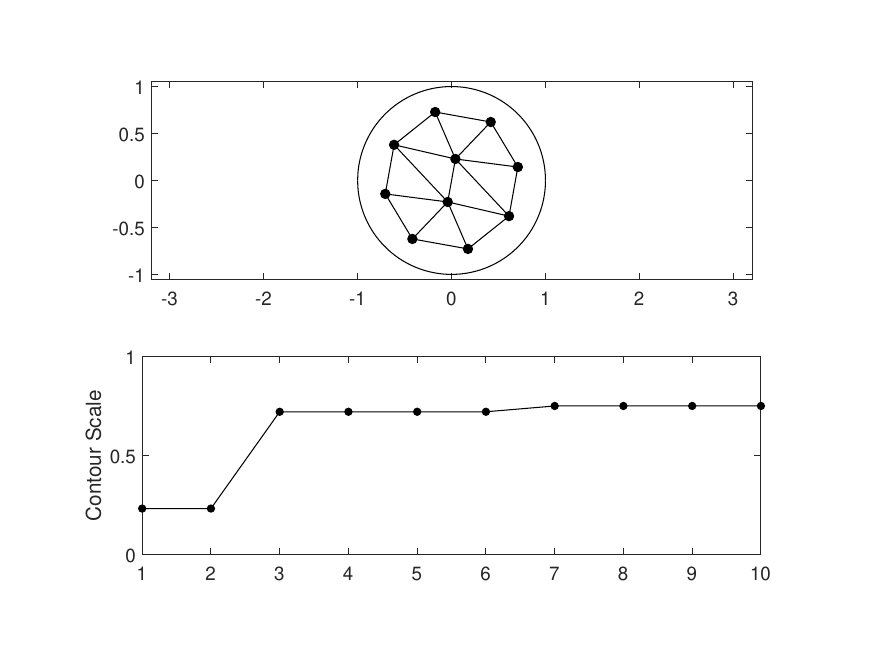}
}
\end{subfigure}
\begin{subfigure}[]{
\includegraphics[width=0.5\textwidth]{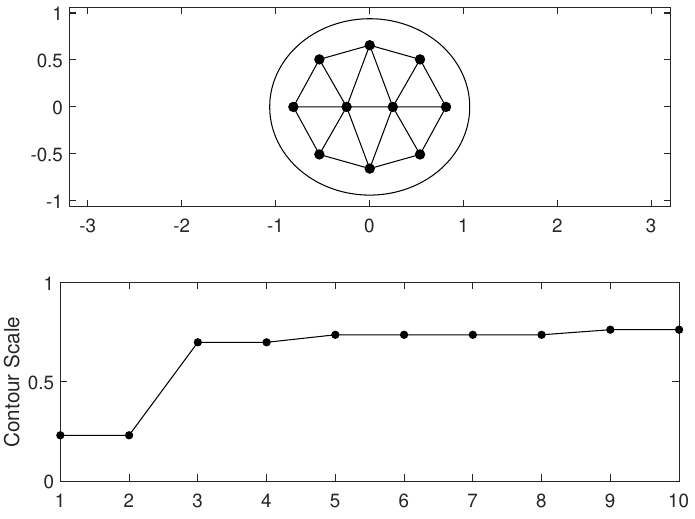}
}
\end{subfigure}
\begin{subfigure}[]{
\includegraphics[width=0.5\textwidth]{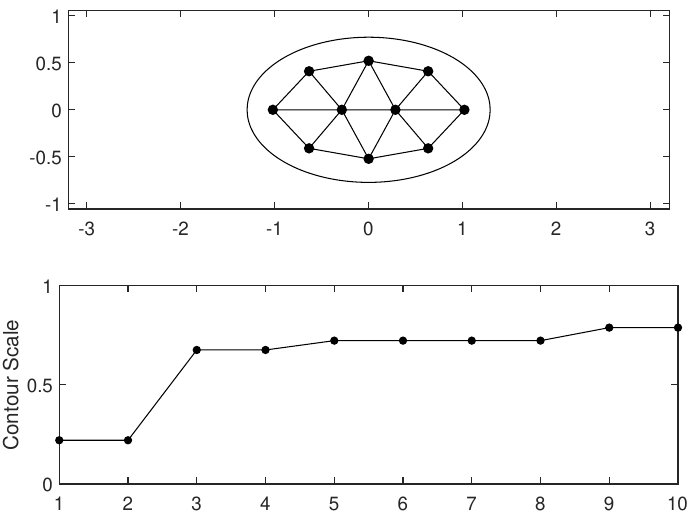}
}
\end{subfigure}
\begin{subfigure}[]{
\includegraphics[width=0.5\textwidth]{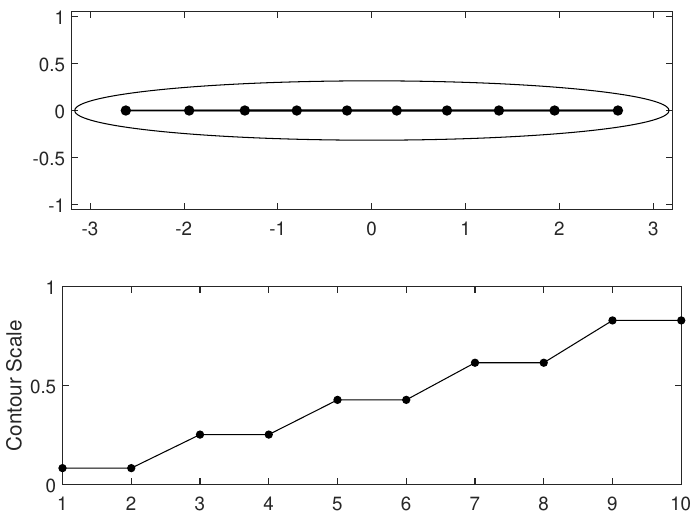}
}
\end{subfigure}
\caption{\small Plots depicting optimal trap distribution for $N=10$, comparing eccentricities of (a) $\ecc = 0$, (b) $\ecc = 0.472$, (c) $\ecc = 0.802$, (d) $\ecc = 0.995$. Upper plots show positions of traps along with a crude visualization of nearest-neighbour pairs calculated using Delaunay triangulation. Lower plots show the scaling factor given by the equation \eqref{eq:scaleFact}.}
\label{fig:configN10}
\end{figure}

\begin{figure}[htbp]
\begin{subfigure}[]{
\includegraphics[width=0.5\textwidth]{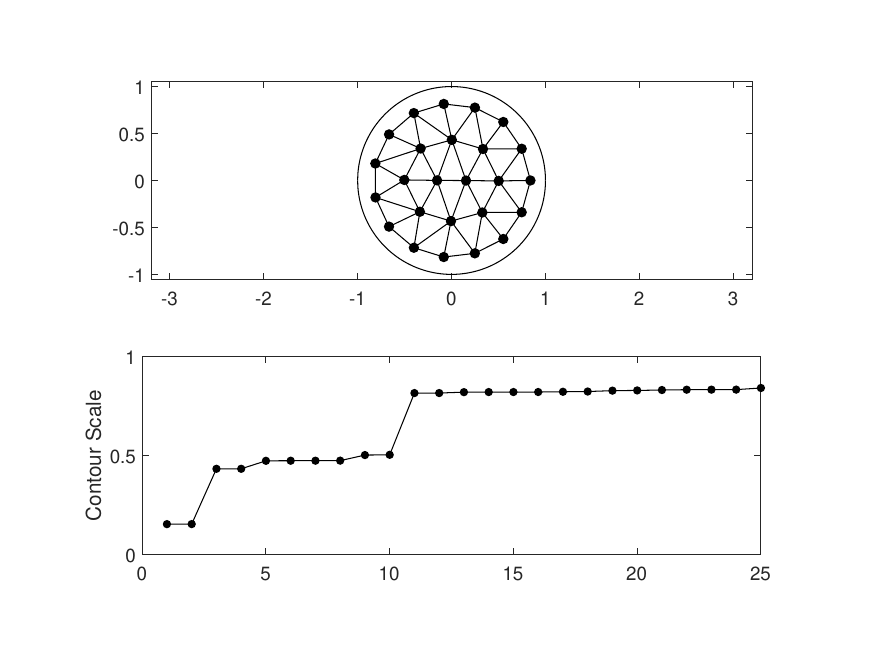}
}
\end{subfigure}
\begin{subfigure}[]{
\includegraphics[width=0.5\textwidth]{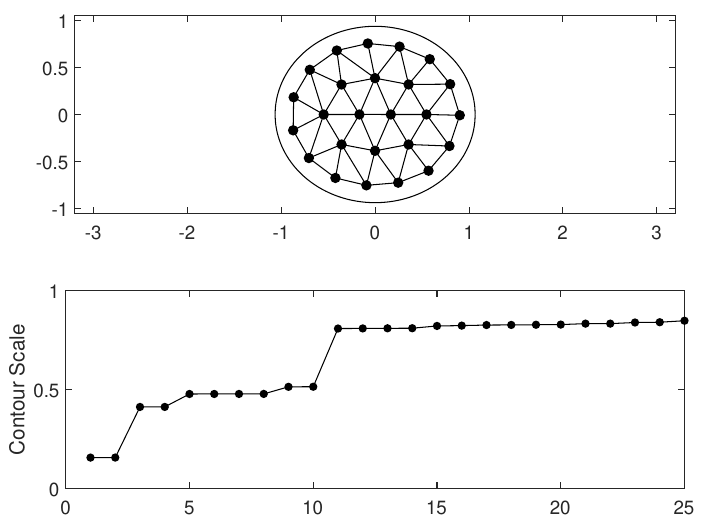}
}
\end{subfigure}
\begin{subfigure}[]{
\includegraphics[width=0.5\textwidth]{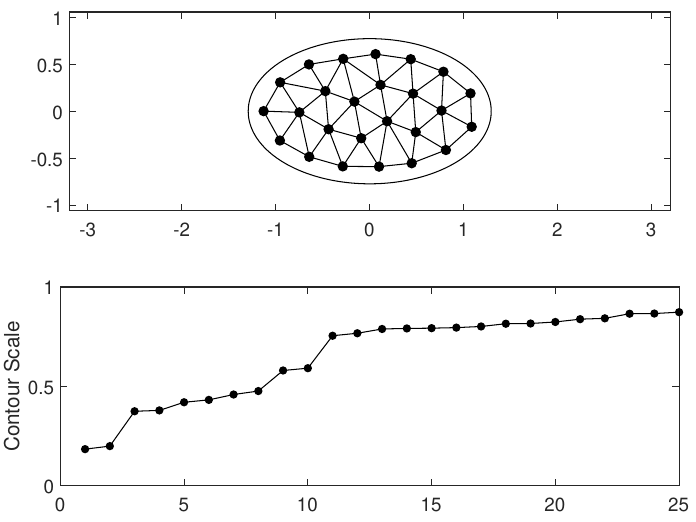}
}
\end{subfigure}
\begin{subfigure}[]{
\includegraphics[width=0.5\textwidth]{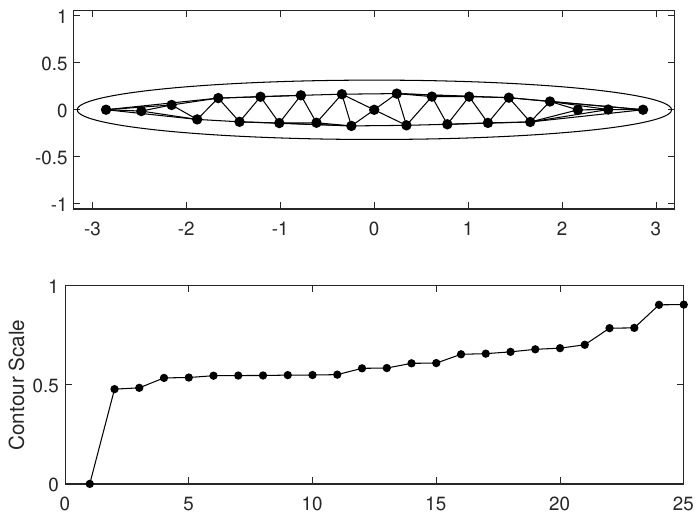}
}
\end{subfigure}
\caption{\small Plots depicting optimal trap distribution for $N=25$, comparing eccentricities of (a) $\ecc = 0$, (b) $\ecc = 0.472$, (c) $\ecc = 0.802$, (d) $\ecc = 0.995$. Upper plots show positions of traps along with a crude visualization of nearest-neighbour pairs calculated using Delaunay triangulation. Lower plots show the scaling factor given by the equation  \eqref{eq:scaleFact}.}
\label{fig:configN25}
\end{figure}

\begin{figure}[htbp]
\begin{subfigure}[]{
\includegraphics[width=0.5\textwidth]{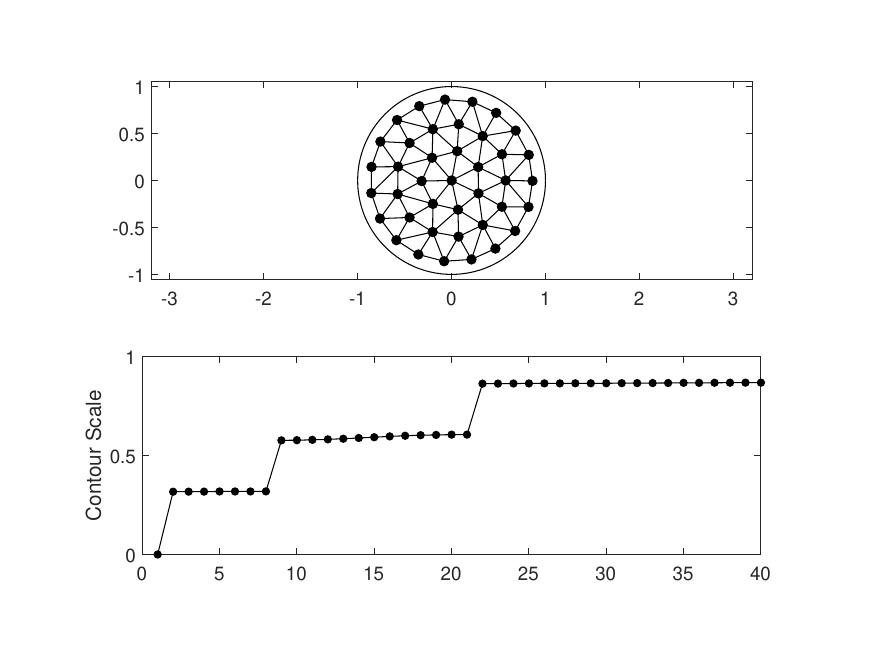}
}
\end{subfigure}
\begin{subfigure}[]{
\includegraphics[width=0.5\textwidth]{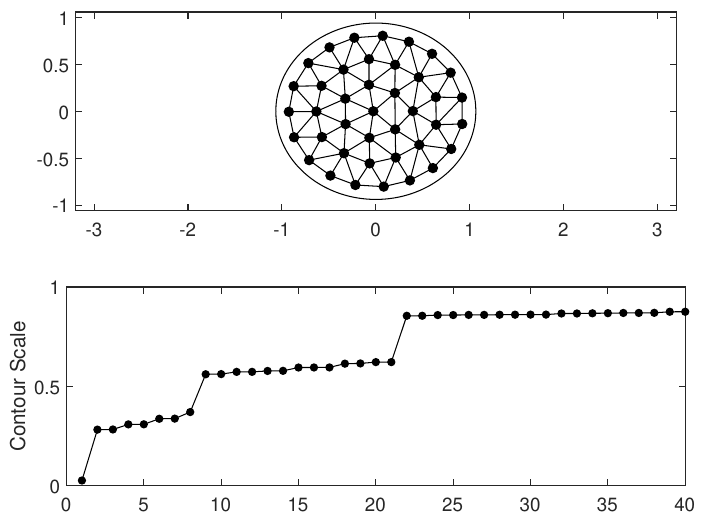}
}
\end{subfigure}
\begin{subfigure}[]{
\includegraphics[width=0.5\textwidth]{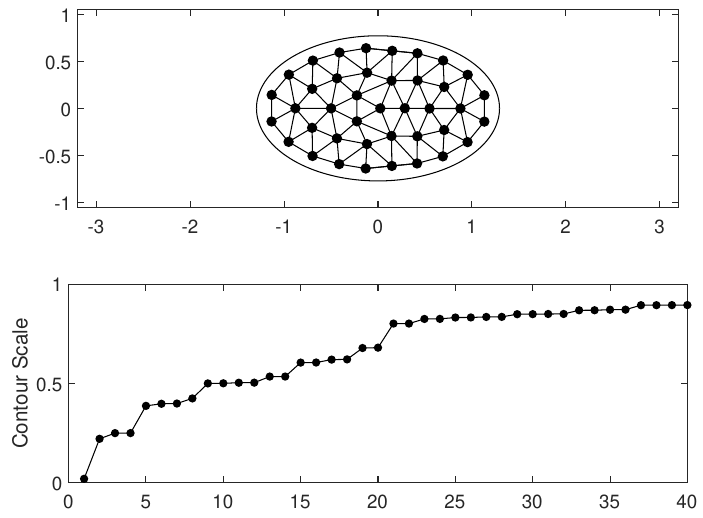}
}
\end{subfigure}
\begin{subfigure}[]{
\includegraphics[width=0.5\textwidth]{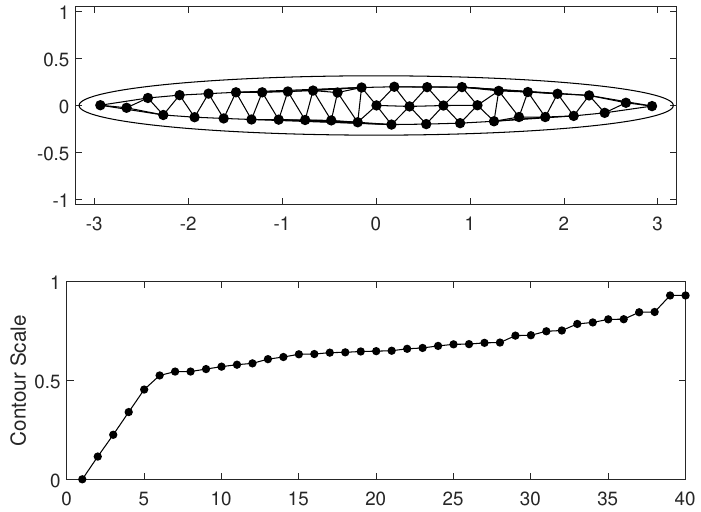}
}
\end{subfigure}
\caption{\small Plots depicting optimal trap distribution for $N=40$, comparing eccentricities of (a) $\ecc = 0$, (b) $\ecc = 0.472$, (c) $\ecc = 0.802$, (d) $\ecc = 0.995$. Upper plots show positions of traps along with a crude visualization of nearest-neighbour pairs calculated using Delaunay triangulation. Lower plots show the scaling factor given by the equation  \eqref{eq:scaleFact}.}
\label{fig:configN40}
\end{figure}

Plots comparing the optimal configurations of select $N$ for each of the eccentricities considered in this study are shown in Figures \ref{fig:configN5}--\ref{fig:configN40}. Each plot shows the positions of the traps within the domain, along with a visualization of a Delaunay triangulation \cite{lee1980two} calculated using the traps as vertices, to illustrate the distribution of the traps and relative distance between them.

In addition, it was of interest to see how the ring-like distribution of traps would change with the eccentricity of the domain. This is related to the observation that for the unit disk, optimal trap configurations tend to form ring-type structures \cite{kolokolnikov2005optimizing}; a similar effect is observed for the MFPT problem with internal traps in the unit sphere, where equal traps tend to be centered close to nested spherical surfaces \cite{gilbert2019globally}. In order to visualize possible ring-like elliptic configurations, a scaling factor was calculated for each trap such that each trap would lie on a scaled copy of the elliptic domain boundary. This scaling factor $c$ is given by the equation
\begin{equation} \label{eq:scaleFact}
\left( \dfrac{x}{a} \right)^2 \ps \left( \dfrac{y}{b} \right)^2 \es c^2 \ ,
\end{equation}
where $c = 1$ corresponds to the boundary of the domain. When the domain is circular, meaning $a = b = 1$, then in terms of polar coordinates, $c$ is the radial coordinate of the trap. These scaling factors are shown in the lower subplots in Figures \ref{fig:configN5}--\ref{fig:configN40}. For the case of $N=5$, Figure \ref{fig:configN5} can be compared to the optimal configurations presented in Ref.~\cite{iyaniwura2021optimization}, through which it can be seen that the two are qualitatively similar and exhibit the same relationship between trap distribution and domain eccentricity. For higher eccentricity values (except the degenerate cases where traps lie on the symmetry axis: Figures \ref{fig:configN5}(d), \ref{fig:configN10}(d)), the trap configurations do not tend to show ring-like structures that are present, to some degree, in the disk and low-eccentricity elliptic domains.

To examine the putative optimal distributions of traps in terms of their pairwise distance to the closest neighbor and related measures, a Delaunay triangulation was computed to identify neighbors of each trap (see upper plots in Figures \ref{fig:configN5}--\ref{fig:configN40}). In general, for a configuration of $N$ traps distributed, in some sense, ``uniformly" over the elliptic domain of area $\domMeas = \pi$, the average ``area per trap" is given by $A(N) = \domMeas/N = \pi/N$. Likening an optimal arrangement of $N$ traps to a collection of circles packed into an enclosed space, the (average)  distance $\langle d\rangle$  between two neighbouring traps would be the distance between the centers of two identical circles representing the area occupied by each trap; it would be related to the area per trap as $A(N)= \pi\langle d\rangle^2 / 4$. One consequently finds that the average  distance between neighbouring traps, equivalent to the diameter of one of the circles, is given by
\begin{equation} \label{eq:avgDist}
\langle d \rangle \es \sqrt{\dfrac{4 \domMeas}{\pi N}} \es \dfrac{2}{\sqrt{N}} \ .
\end{equation}
Extending this comparison to the traps nearest the boundary, the smallest distance between a trap and the boundary was taken to be the radius of a circle surrounding the trap, and the diameter of this circle was compared to the smallest distance between two traps. This essentially provides a measure of the distance between a near-the-boundary trap and its ``reflection" in the Neumann boundary.

\begin{figure}[H]
\begin{subfigure}[]{
\includegraphics[width=0.5\textwidth]{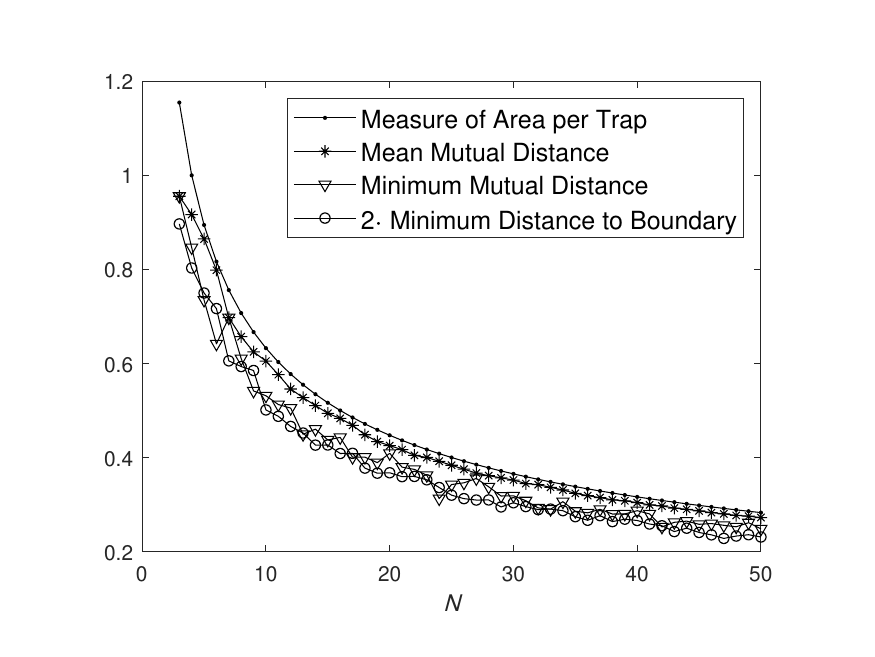}
}
\end{subfigure}
\begin{subfigure}[]{
\includegraphics[width=0.5\textwidth]{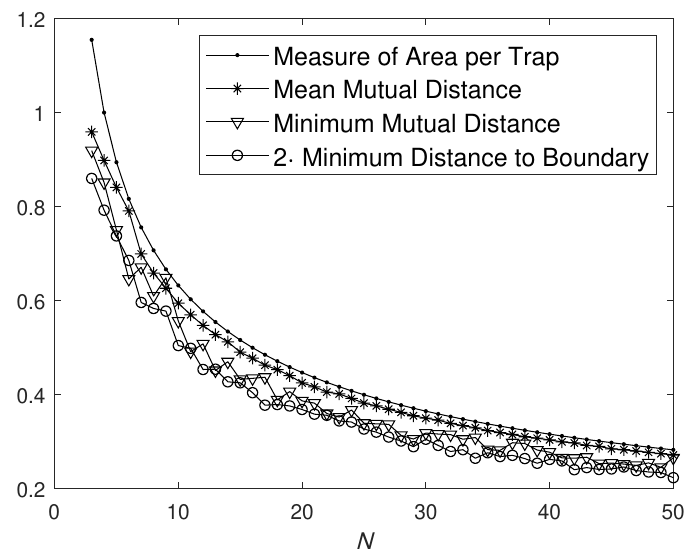}
}
\end{subfigure}
\begin{subfigure}[]{
\includegraphics[width=0.5\textwidth]{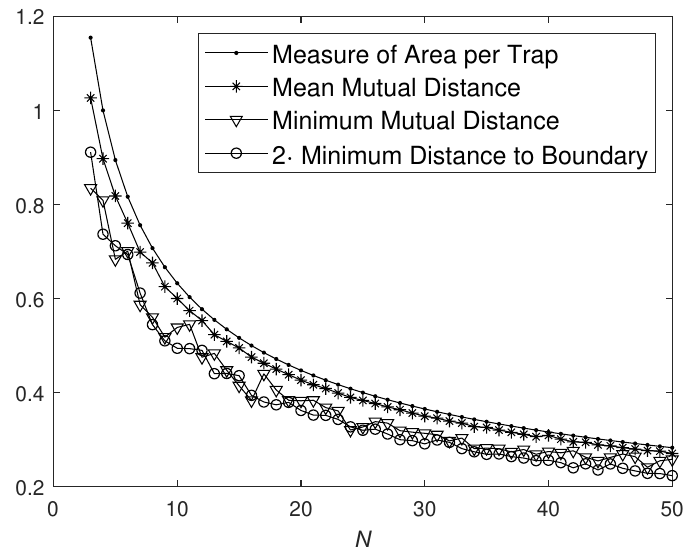}
}
\end{subfigure}
\begin{subfigure}[]{
\includegraphics[width=0.5\textwidth]{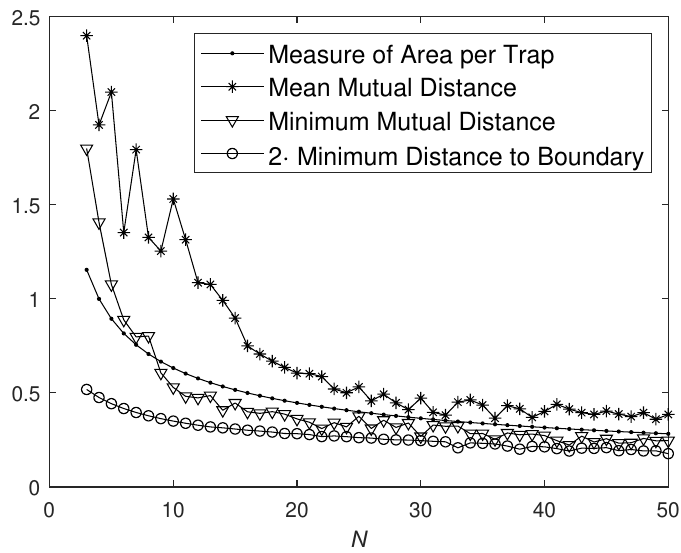}
}
\end{subfigure}
\caption{\small Plots depicting local pairwise distance properties of optimal trap distributions as functions of $N$, for domain eccentricities  $\ecc = 0$ (a), $\ecc = 0.472$ (b), $\ecc = 0.802$ (c), and $\ecc = 0.995$ (d). The curve entitled ``Measure of Area per Trap" shows the distance $\langle d \rangle$ computed using the ``area per trap" argument and the resulting formula \eqref{eq:avgDist}.
}
\label{fig:distComp}
\end{figure}

In Figure \ref{fig:distComp}, for each of the four considered eccentricities of the elliptic domain, the mean pairwise distance between neighbouring traps is plotted as a function of $N$, along with minimum pairwise distance between traps, and $2\times$ minimal distance to the boundary. These are compared with the average distance formula \eqref{eq:avgDist} coming from the ``area per trap" argument. It can be observed that the simple formula \eqref{eq:avgDist} may be used as a reasonable estimate of common pairwise distances between traps in an optimal configuration.

\section{Traps of different sizes}\label{sec:traps:different}


We now consider the case when the small traps are circular and well-separated but have differing sizes given by $\veps_j$, $j=1,\ldots, N$, with the corresponding size parameters $\nu_j = -1/\log\veps_j$. In the case of same or different-sized traps, the leading-term MFPT contribution satisfies \cite{iyaniwura2021optimization}
\[
u_0(\trapLoc)=-2\pi\sum_{k=1}^{N}\vecA_k G(\trapLoc, \trapLoc_0) + \bar{u}_0
\]
and, when matched with the far-field behaviour of $u_0$, yields
\begin{equation}\label{eq:2:13}
\bar{u}_0 - 2\pi \left(\vecA_j  R_j + \sum_{i \neq j}^{N} \vecA_j G(\trapLoc_j ; \trapLoc_i)\right)\sim \dfrac{\vecA_j}{\nu_j}\,, \qquad j=1,\ldots, N.
\end{equation}
When all $\nu_j=\nu$, the matching condition \eqref{eq:2:13} and \eqref{eq:ellAMFPT:A:sum} yield the formulas \eqref{eq:ellAMFPT} and \eqref{eq:ellAMFPT:A}. In the general case, the equations \eqref{eq:2:13} and \eqref{eq:ellAMFPT:A:sum} can also be solved explicitly. One obtains
\begin{equation}\label{eq:u0:sol:noneq}
\bar{u}_0=\dfrac{\domMeas}{2\pi D \bar{\nu} N} + \dfrac{1}{\bar{\nu} N}\,\vec{w}\cdot \vecA\,,
\end{equation}
where the vector $\vecA$ is the solution of the linear system
\begin{equation}\label{eq:A:sol:noneq}
(I+2\pi \mathcal{N} \greenMat-\mathcal{Q})\,\vecA=\dfrac{\domMeas}{2\pi D \bar{\nu} N} \,\vec{\nu}.
\end{equation}
In \eqref{eq:u0:sol:noneq} and \eqref{eq:A:sol:noneq}, $\greenMat$ is the Green's matrix \eqref{eq:ellGmat}, $\vec{\nu}=(\nu_1,\ldots, \nu_N)^T$ is the trap size parameter vector, $\mathcal{N}={\rm diag}\,\vec{\nu}$ is the corresponding diagonal $N\times N$ matrix,  $\bar{\nu}=(\sum_{i=1}^{N}\,\nu_j)/N$ is the average trap size parameter, $\vec{w} = 2\pi \vec{e}^T\mathcal{N}\greenMat$ is a row vector, and $\mathcal{Q}=(\vec{\nu}\cdot \vec{w})/(\bar{\nu} N)$ is an $N\times N$ matrix.

The optimization of the approximate AMFPT \eqref{eq:u0:sol:noneq} corresponds to the optimization of the second term of \eqref{eq:u0:sol:noneq}, or equivalently, the merit function
\begin{equation} \label{eq:merit:noneq:traps}
\meritFunc(\trapLoc_1,\ldots,\trapLoc_N) \es \vec{e}^T\mathcal{N}\greenMat \vecA \
\end{equation}
depending on $2N$ scalar variables. The expression \eqref{eq:merit:noneq:traps} generalizes the merit function \eqref{eq:ellMerit} onto the setup with nonequal traps.

\begin{figure}[H]
\begin{subfigure}[]{
\includegraphics[width=0.5\textwidth]{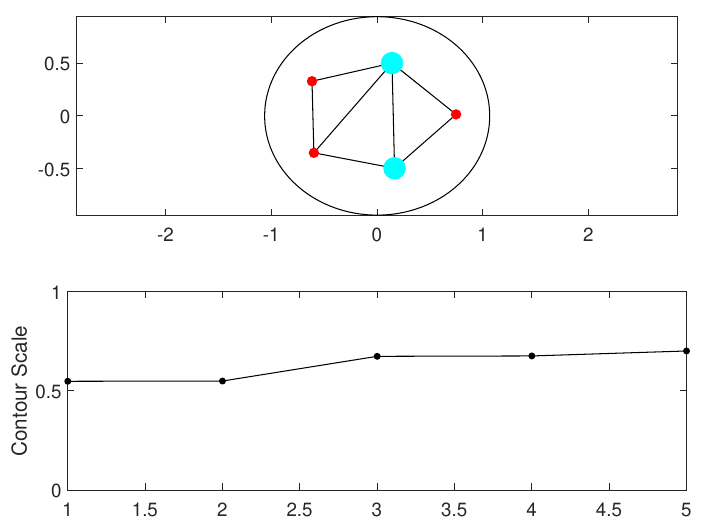}
}
\end{subfigure}
\begin{subfigure}[]{
\includegraphics[width=0.5\textwidth]{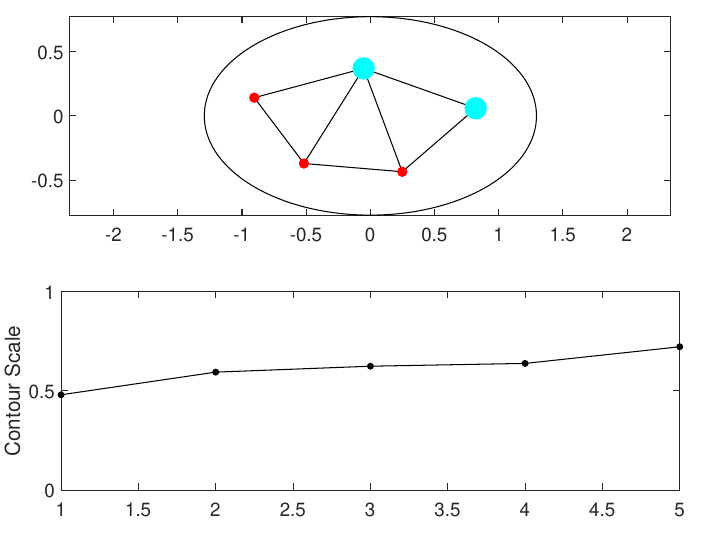}
}
\end{subfigure}
\begin{subfigure}[]{
\includegraphics[width=0.5\textwidth]{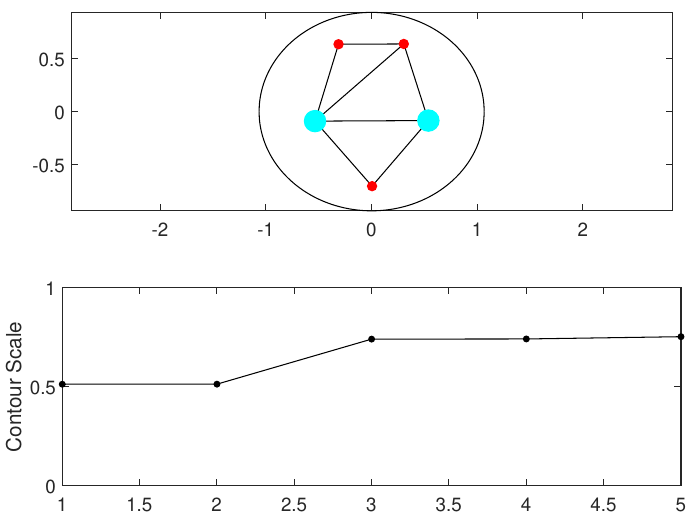}
}
\end{subfigure}
\begin{subfigure}[]{
\includegraphics[width=0.5\textwidth]{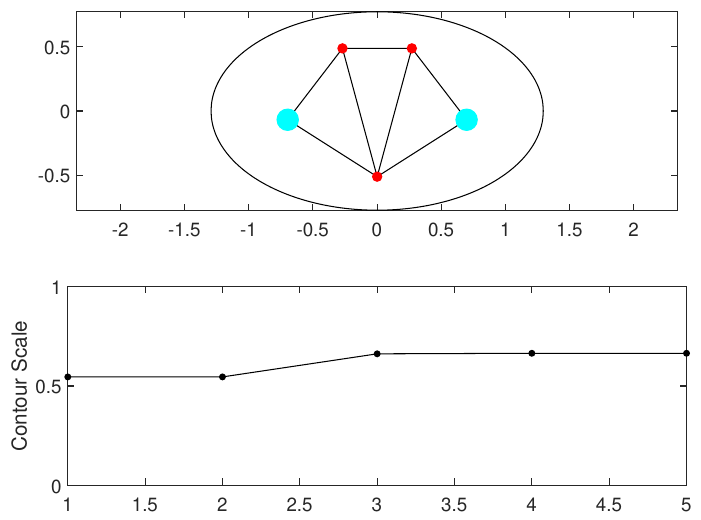}
}
\end{subfigure}
\caption{\small Plots depicting optimal trap distributions for $N=5$ traps with three traps of radius $\veps_1 = 10^{-9}$ and two larger traps of radius $\veps_2 = 10^{3}\veps_1$ (upper) or $\veps_2 = 10^{6}\veps_1$ (lower), and $\ecc = 0.472$ (left) or $\ecc = 0.802$ (right).  Upper plots show positions of traps along with a crude visualization of nearest-neighbour pairs calculated using Delaunay triangulation. Lower plots show the scaling factor given by equation \eqref{eq:scaleFact}.}
\label{fig:configDiffN5}
\end{figure}
\begin{figure}[H]
\begin{subfigure}[]{
\includegraphics[width=0.5\textwidth]{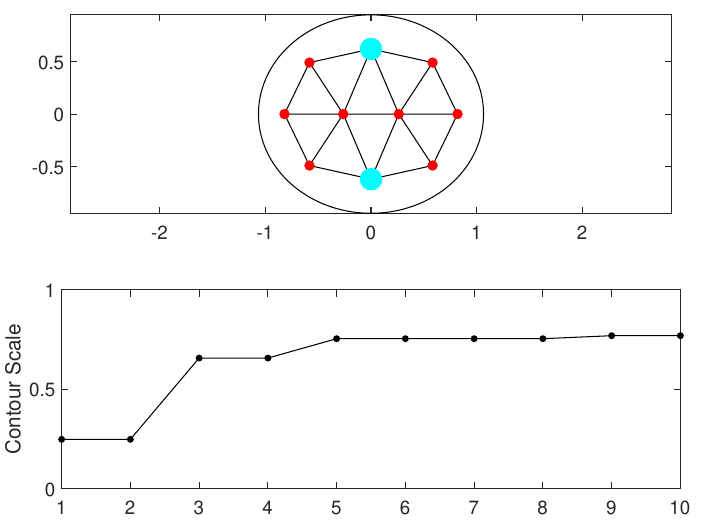}
}
\end{subfigure}
\begin{subfigure}[]{
\includegraphics[width=0.5\textwidth]{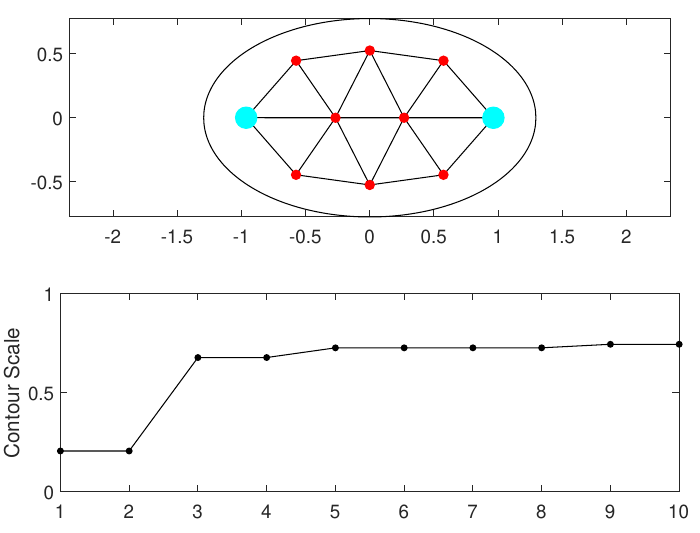}
}
\end{subfigure}
\begin{subfigure}[]{
\includegraphics[width=0.5\textwidth]{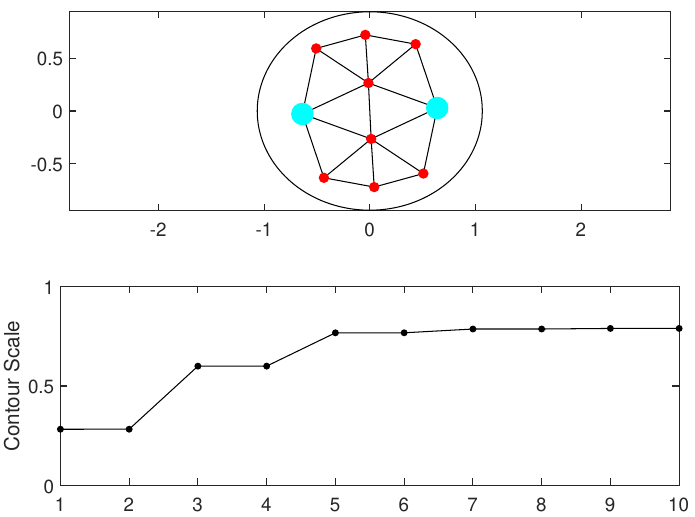}
}
\end{subfigure}
\begin{subfigure}[]{
\includegraphics[width=0.5\textwidth]{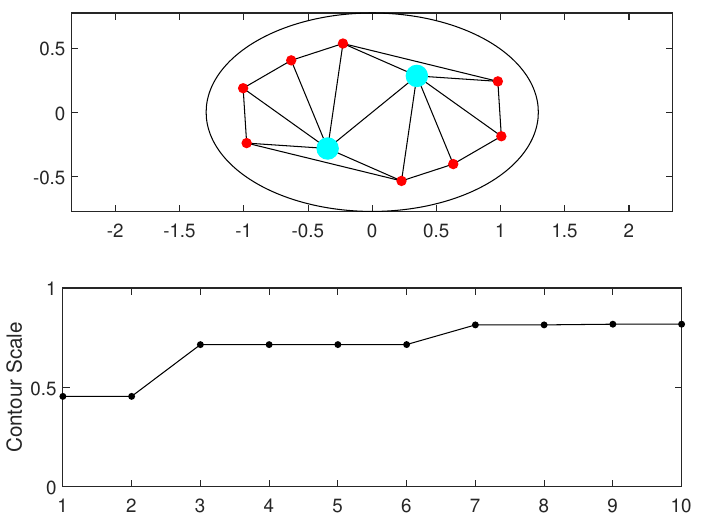}
}
\end{subfigure}
\caption{\small Plots depicting optimal trap distributions for $N=10$ traps with eight traps of radius $\veps_1 = 10^{-9}$ and two larger traps of radius $\veps_2 = 10^{3}\veps_1$ (upper) or $\veps_2 = 10^{6}\veps_1$ (lower), and $\ecc = 0.472$ (left) or $\ecc = 0.802$ (right).  Upper plots show positions of traps along with a crude visualization of nearest-neighbour pairs calculated using Delaunay triangulation. Lower plots show the scaling factor given by equation \eqref{eq:scaleFact}.}
\label{fig:configDiffN10}
\end{figure}

\medskip As an illustration, we consider elliptic domains with $N=N_1+N_2$ traps, such that $N_1$ traps have common radii $\veps_1$, and $N_2$ traps common radii $\veps_2>\veps_1$. Using the global optimization procedure described in Section \ref{sec:method} and the values $\veps_1 = 10^{-9}$ and $\veps_2 = 10^{3}\veps_1$ or $\veps_2 = 10^{6}\veps_1$, the configurations obtained for two different values of eccentricity  $\ecc = 0.472$ or $\ecc = 0.802$ and the trap numbers $N_2=2$ and $N=5, 10$ are shown in Figures \ref{fig:configDiffN5} and \ref{fig:configDiffN10}. As expected, larger traps produce a stronger ``push" than the smaller ones. It is evident that the increase of the $\veps_2/\veps_1$ ratio causes a significant change in the form of the configuration.

\section{Discussion} \label{sec:discussion}

At this point some interpretation of the previously stated results will be presented. This discussion will concern the putative optimal values of the average MFPT \eqref{eq:ellAMFPT} in elliptic domains with internal traps, values of the related merit function \eqref{eq:ellMerit}, the positions of traps within the domain, the bulk measures of trap distribution which were employed, and the configurations of non-identical traps.

To begin, the method of study will be briefly reiterated. In order to study the dependence of optimal trap configurations on both the number of traps, and the eccentricity of the elliptic domain, the values of the merit function \eqref{eq:ellMerit} corresponding to the approximate average MFPT \eqref{eq:amfpt:new} were minimized for $N \leq 50$ and sample values of the eccentricity \eqref{eq:eccDef} of 0, 0.472, 0.802, and 0.995, while the area of the ellipse was kept constant, $\domMeas = \pi$. For $N$ traps, the merit function consequently depends on $2N$ scalar variables. In the search for a global optimum, an iterative approach, which switched between global and local searches, was used (Section \ref{sec:method}). The method used here was similar to that of Ref.~\cite{iyaniwura2021optimization}, though a different algorithm was used for the local search, as well as in Ref.~\cite{gilbert2019globally} which used a different algorithms than here for both searches. A somewhat different approach was employed in Ref.~\cite{iyaniwura2019simulation}, which made use of numerical solutions to the Poisson problem \eqref{eq:defPDE}. In the case of the unit disk, the comparison of the results of this study to those in Ref.~\cite{kolokolnikov2005optimizing} demonstrated that the optima reported here are consistently an improvement on previous work. In particular, this is due to the removal of the constraint that all traps be located on rings within the domain -- see, e.g., Figure \ref{fig:configN10}(a) where for $N=10$, trap centers evidently do not lie on concentric rings.

Plots of the putative globally optimal merit function values vs.~$N$, for each eccentricity value considered, are found in Figure \ref{fig:meritComp}. It follows that the domain eccentricity is a more important factor when there are few traps, but each function behaves similarly as $N$ increases. In particular, as the number of traps $N$ increases, the average MFPT $\amfpt$ \eqref{eq:ellAMFPT} approaches zero; the merit function $\meritFunc(\trapLoc)$ therefore must, as $N\to \infty$, approach from above the value $-\domMeas / (4\pi^2 D\nu) \simeq -0.238$, which agrees with the plots in Figure \ref{fig:meritComp}.

Examination of the positions of traps in the AMFPT-minimizing configurations, both visually and in terms of their radial coordinates, gives the impression that the optimal configuration is one which consists of traps placed on the vertices of nested polygons. These polygons, while irregular, seem to possess some consistent structure, including being convex. (It is interesting to note that the optimal configurations of confined interacting points often take similar forms, both in two and three dimensions \cite{sloane1995minimal, manoharan2003dense, saint2001macroscopic, hoare1971physical}.) Due to the geometrical symmetries of the ellipse, the optimal configurations are defined uniquely modulo the group $C^2\times C^2$ of reflections with respect to both axes, which includes the rotation by $\pi$. Numerical optimization algorithms converge to a single specific representative of the equivalent putative globally optimal configurations. For example, for non-symmetric numerically optimal configuration, several traps may be found along the midline of one half (right or left) of the domain. Optimal trap configurations with the same symmetry group as the ellipse were also observed (see, e.g., Figure~\ref{fig:configN10}(c)).

In addition to the examination of individual trap positions in each optimized configuration, quantities were calculated using the distances between neighbouring traps, defined using a Delaunay triangulation of the trap coordinates, which served as bulk measures of the distribution of traps in each configuration. Plots of these measures, shown in Figure \ref{fig:distComp}, illustrate that the mean distance between neighbouring traps tends to be close to the diameter of a circle which would occupy the average area of the domain per trap, as in equation \eqref{eq:avgDist}. Additionally, the minimum distance between any two traps tends to be twice the minimum distance between a trap and the domain boundary, which supports the intuitive reasoning that for the boundary value problem \eqref{eq:defPDE} with interior traps, the Neumann boundary condition on $\partial\Omega$ ``reflects" each trap, so that under the average MFPT optimization, every trap tends to ``repel" from its reflection in the boundary the same way as it is repelled from other traps.

In Section \ref{sec:traps:different}, the formulas obtained in Ref.~\cite{iyaniwura2019simulation} pertaining to the approximate asymptotic MFPT in an elliptic domain were generalized onto the case of $N$ traps of different sizes, defined by the radii $\veps_j$, $j=1,\ldots, N$. It was shown that that the AMFPT is approximated by \eqref{eq:u0:sol:noneq}, and can be minimized simultaneously with the merit function \eqref{eq:merit:noneq:traps} of $2N$ scalar trap coordinates. Global optimization was performed for sample configurations corresponding to all combinations of trap numbers $N = \lbrace 5, 10 \rbrace$, domain eccentricities $\ecc = \lbrace 0.472, 0.802 \rbrace$, and trap size relations $\veps_2 = \lbrace 10^{3}\veps_1, 10^{6}\veps_1 \rbrace$, when two traps had the same radius $\veps_2$ and $N-2$ traps the radius $\veps_1$. The respective putative optimal configurations shown in Figures \ref{fig:configDiffN5}--\ref{fig:configDiffN10} were obtained. In particular, strong dependence of the form of the optimal configuration on the trap size ratio was observed.

Both in the case of same and different trap sizes, multiple $N$-trap configurations corresponding to local minima of the AMFPT exist, some having rather close values of the merit function. Computation and analysis of the structure of such local minima, possibly along the lines of a similar study for traps on the surface of the sphere \cite{ridgway2019locally}, is a possible direction of future research.

In future work, it would also be of interest to address the following related problems. The first is to carry out similar investigations for non-elliptic near-disk domains considered in Ref.~\cite{iyaniwura2021optimization}. Another interesting direction is the development of a scaling law which would predict the behaviour of the MFPT as the number of traps increases with their positions defined according to a specific distribution, in particular, for distributions that globally or locally minimize MFPT, or other distributions of practical significance. A similar problem, along with the dilute trap fraction limit of homogenization theory, was addressed in Refs.~\cite{cheviakov2010asymptotic, cheviakov2013narrow} for the narrow escape problem involving boundary traps located on the surface of a sphere in three dimensions.

\subsection*{Acknowledgements}
A.C. is thankful to NSERC of Canada for research support through the Discovery grant RGPIN-2019-05570.

\footnotesize
\bibliography{NarrowCaptureEllipseOpt}

\begin{thebibliography}{10}

\bibitem{iyaniwura2021optimization}
Sarafa~A Iyaniwura, Tony Wong, Colin~B Macdonald, and Michael~J Ward.
\newblock Optimization of the mean first passage time in near-disk and
  elliptical domains in {2-D} with small absorbing traps.
\newblock {\em SIAM Review}, 63(3):525--555, 2021.

\bibitem{red2001}
Sidney Redner.
\newblock {\em A Guide to First-Passage Processes}.
\newblock Cambridge University Press, 2001.

\bibitem{holcman2014narrow}
David Holcman and Zeev Schuss.
\newblock The narrow escape problem.
\newblock {\em SIAM Review}, 56(2):213--257, 2014.

\bibitem{saffman1975brownian}
PG~Saffman and M~Delbr{\"u}ck.
\newblock Brownian motion in biological membranes.
\newblock {\em Proceedings of the National Academy of Sciences},
  72(8):3111--3113, 1975.

\bibitem{ralf2014first}
Ralf Metzler, Sidney Redner, and Gleb Oshanin.
\newblock {\em First-passage Phenomena and their Applications}, volume~35.
\newblock World Scientific, 2014.

\bibitem{bressloff2008diffusion}
Paul~C Bressloff, Berton~A Earnshaw, and Michael~J Ward.
\newblock Diffusion of protein receptors on a cylindrical dendritic membrane
  with partially absorbing traps.
\newblock {\em SIAM Journal on Applied Mathematics}, 68(5):1223--1246, 2008.

\bibitem{bressloff2013stochastic}
Paul~C Bressloff and Jay~M Newby.
\newblock Stochastic models of intracellular transport.
\newblock {\em Reviews of Modern Physics}, 85(1):135, 2013.

\bibitem{holcman2004escape}
D~Holcman and Z~Schuss.
\newblock Escape through a small opening: receptor trafficking in a synaptic
  membrane.
\newblock {\em Journal of Statistical Physics}, 117(5-6):975--1014, 2004.

\bibitem{kolokolnikov2005optimizing}
Theodore Kolokolnikov, Michele~S Titcombe, and Michael~J Ward.
\newblock Optimizing the fundamental {N}eumann eigenvalue for the {L}aplacian
  in a domain with small traps.
\newblock {\em European Journal of Applied Mathematics}, 16(2):161--200, 2005.

\bibitem{cheviakov2011optimizing}
Alexei~F Cheviakov and Michael~J Ward.
\newblock Optimizing the principal eigenvalue of the {L}aplacian in a sphere
  with interior traps.
\newblock {\em Mathematical and Computer Modelling}, 53(7-8):1394--1409, 2011.

\bibitem{gilbert2019globally}
Jason Gilbert and Alexei Cheviakov.
\newblock Globally optimal volume-trap arrangements for the narrow-capture
  problem inside a unit sphere.
\newblock {\em Physical Review E}, 99(1):012109, 2019.

\bibitem{ridgway2019locally}
Wesley~JM Ridgway and Alexei~F Cheviakov.
\newblock Locally and globally optimal configurations of {N} particles on the
  sphere with applications in the narrow escape and narrow capture problems.
\newblock {\em Physical Review E}, 100(4):042413, 2019.

\bibitem{vaz2007particle}
A~Ismael~F Vaz and Lu{\'\i}s~N Vicente.
\newblock A particle swarm pattern search method for bound constrained global
  optimization.
\newblock {\em Journal of Global Optimization}, 39(2):197--219, 2007.

\bibitem{currie2012opti}
Jonathan Currie, David~I Wilson, Nick Sahinidis, and Jose Pinto.
\newblock {OPTI}: {L}owering the barrier between open source optimizers and the
  industrial {MATLAB} user.
\newblock {\em Foundations of computer-aided process operations}, 24:32, 2012.

\bibitem{lagarias1998convergence}
Jeffrey~C Lagarias, James~A Reeds, Margaret~H Wright, and Paul~E Wright.
\newblock Convergence properties of the {N}elder--{M}ead simplex method in low
  dimensions.
\newblock {\em SIAM Journal on optimization}, 9(1):112--147, 1998.

\bibitem{lee1980two}
Der-Tsai Lee and Bruce~J Schachter.
\newblock Two algorithms for constructing a {D}elaunay triangulation.
\newblock {\em International Journal of Computer \& Information Sciences},
  9(3):219--242, 1980.

\bibitem{iyaniwura2019simulation}
Sarafa Iyaniwura, Tony Wong, Michael~J Ward, and Colin~B Macdonald.
\newblock Simulation and optimization of mean first passage time problems in
  2-{D} using numerical embedded methods and perturbation theory.
\newblock {\em arXiv preprint arXiv:1911.07842}, 2019.

\bibitem{sloane1995minimal}
Neil~JA Sloane, Ronald~H Hardin, TDS Duff, and John~H Conway.
\newblock Minimal-energy clusters of hard spheres.
\newblock {\em Discrete \& Computational Geometry}, 14(3):237--259, 1995.

\bibitem{manoharan2003dense}
Vinothan~N Manoharan, Mark~T Elsesser, and David~J Pine.
\newblock Dense packing and symmetry in small clusters of microspheres.
\newblock {\em Science}, 301(5632):483--487, 2003.

\bibitem{saint2001macroscopic}
M~Saint~Jean, C~Even, and C~Guthmann.
\newblock Macroscopic 2{D} {W}igner islands.
\newblock {\em EPL (Europhysics Letters)}, 55(1):45, 2001.

\bibitem{hoare1971physical}
MR~Hoare and P~Pal.
\newblock Physical cluster mechanics: Statics and energy surfaces for monatomic
  systems.
\newblock {\em Advances in Physics}, 20(84):161--196, 1971.

\bibitem{cheviakov2010asymptotic}
Alexei~F Cheviakov, Michael~J Ward, and Ronny Straube.
\newblock An asymptotic analysis of the mean first passage time for narrow
  escape problems: {P}art {II:} the sphere.
\newblock {\em Multiscale Modeling \& Simulation}, 8(3):836--870, 2010.

\bibitem{cheviakov2013narrow}
Alexei~F Cheviakov and Daniel Zawada.
\newblock Narrow-escape problem for the unit sphere: Homogenization limit,
  optimal arrangements of large numbers of traps, and the {N}-square
  conjecture.
\newblock {\em Physical Review E}, 87(4):042118, 2013.

\end{thebibliography}
\bibliographystyle{unsrt}

\end{document}